\documentclass[aps,prl,reprint,superscriptaddress,amsmath,amssymb,floatfix]{revtex4-2}
\usepackage[T1]{fontenc}
\usepackage{graphicx}
\usepackage{bm}
\usepackage[colorlinks=true,citecolor=blue,linkcolor=blue,urlcolor=blue]{hyperref}
\graphicspath{{figures/}}
\newcommand{\NR}{N_{\mathbb R}}
\newcommand{\Ec}{E_c}
\newcommand{\cD}{\mathcal D}
\newcommand{\dd}{\delta}

\usepackage{booktabs}
\usepackage{ragged2e}

\begin{document}
\title{\texorpdfstring{Cooperative Boundary Switching\\
of Mesoscopic Real Spectra in Non-Hermitian Chains}
{Cooperative Boundary Switching of Mesoscopic Real Spectra in Non-Hermitian Chains}}
\author{Chengxi Li}
\affiliation{Beijing Institute of Petrochemical Technology, Beijing 102617, China}
\affiliation{Beijing National Laboratory for Condensed Matter Physics, Institute of Physics, Chinese Academy of Sciences, Beijing 100190, China}
\author{Hao Zhu}
\email[Corresponding author: ]{haozhu4012@163.com}
\affiliation{Beijing Institute of Petrochemical Technology, Beijing 102617, China}
\affiliation{Beijing National Laboratory for Condensed Matter Physics, Institute of Physics, Chinese Academy of Sciences, Beijing 100190, China}
\author{Tie-Fu Zhang}
\affiliation{Beijing National Laboratory for Condensed Matter Physics, Institute of Physics, Chinese Academy of Sciences, Beijing 100190, China}
\affiliation{Faculty of Science, Kunming University of Science and Technology, Kunming 650500, China}
\author{Wanzi Sun}
\affiliation{Beijing National Laboratory for Condensed Matter Physics, Institute of Physics, Chinese Academy of Sciences, Beijing 100190, China}
\affiliation{Jining University, Qufu, Shandong 273155, China}
\begin{abstract}
Real-energy eigenmodes oscillate without individual exponential growth or decay.
At fixed bulk criticality, we identify an open family of non-Hermitian chains
in which every fixed nonsingular real single-end potential preserves a growing
local population, whereas an open region of two-end potentials removes it.
Single-end control cannot reduce the relative boundary-matching amplitude below
the real-mode threshold; two endpoint reductions multiply and can cross it.
The population grows as the square root of chain length while its spectral
fraction vanishes, despite unchanged local limiting bands.
For a representative chain's four zero-detuning settings, an independent
passive Hermitian double chain converts this switch into resolvable reflection
resonances at one fixed local port under a common coupling scan at fixed probe
carrier. Absolute reflection amplitude contrast and stability margin both
decrease inversely with length. Boundary amplitudes therefore determine a
minimum control requirement that local limiting bands do not specify.
\end{abstract}
\maketitle

Real-energy eigenmodes oscillate without individual exponential growth
or decay, making their boundary control a basic question in non-Hermitian
physics \cite{AshidaGongUeda2020,BergholtzEtAl2021,ElGanainyEtAl2018}.
Non-Bloch band theory describes limiting open-boundary spectra through
complex wave numbers
\cite{YaoWang2018,YokomizoMurakami2019,LeeThomale2019,YokomizoMurakami2020}.
Skin accumulation is closely tied to point-gap topology
\cite{OkumaEtAl2020,ZhangYangFang2020}, with anomalous exceptions
\cite{GuoEtAl2023Anomalous}. Boundary modes have motivated generalized
bulk-boundary correspondence
\cite{Lee2016,KunstEtAl2018,BorgniaEtAl2020,SongYaoWang2019RealSpace}.
Boundary conditions can strongly reorganize finite spectra and their
real-to-complex transitions
\cite{GuoEtAl2021,EdvardssonArdonne2022,LiEtAl2020,YokomizoMurakami2021,BaoEtAl2024,WangChen2025,HuEtAl2024,ChenZhang2023,Longhi2019}.
Control of non-Hermitian response also extends to magneto-optical systems
\cite{RuanEtAl2025MagnetoOptical} and cavity-magnomechanical entanglement
\cite{YuLiuYasir2025Magnomechanics}.
These developments raise a physical question: how many spatially separated
boundary controls are needed to remove a chosen population of real modes
while keeping the bulk fixed?

The difficulty is that these spectral descriptions do not specify the
minimum boundary control. A growing population of real levels can have
vanishing normalized spectral weight, leaving its finite-chain count
unspecified by limiting bands or boundary-independent thermodynamic
spectral moments \cite{ChengEtAl2024}.
Existing finite-chain methods connect spatial propagation to boundary
conditions \cite{BottcherGrudsky2005,DuitsKuijlaars2008,KoekenbierSchulzBaldes2024}
and characterize spectral accumulation \cite{BrownOtto2020};
finite-size spectral scaling has also been studied \cite{QinEtAl2023}.
Determining whether one endpoint suffices requires the relative amplitudes
with which competing waves satisfy both boundaries.
Local observability presents a further obstacle: real eigenvalues alone
guarantee neither visible port weight nor separation from background.
Hermitian doubling
\cite{GongEtAl2018,KawabataEtAl2019Symmetry,PorrasFernandezLorenzo2019}
supplies an independent setting in which these readout requirements can
be assessed.

In this Letter, we establish cooperative boundary switching in an open
analytic family of non-Hermitian chains. At fixed bulk criticality,
every fixed nonsingular real single-end potential preserves a local
population growing as $\sqrt N$, where $N$ is chain length;
an open region of two-end potentials removes it.
We call this growing subextensive population \emph{mesoscopic reality}.
Each endpoint can reduce the boundary-matching amplitude only by a bounded
factor, while the two reductions multiply. This fixes the minimum number
of controlled ends despite unchanged local limiting bands.
For a representative chain's four zero-detuning settings, an independent
passive Hermitian double chain converts the distinction into resolvable
resonance groups at one fixed local port. The settings share a coupling
scan at fixed probe carrier, with explicit signal, stability, and resource
bounds. Contrast and stability margin both decrease inversely with length.

\emph{Model and boundary selection.---}
Consider a noninteracting open chain of $N$ sites with unit lattice
spacing. Its second-quantized Hamiltonian and bulk band function are
\begin{figure}[b]
\includegraphics[width=\columnwidth]{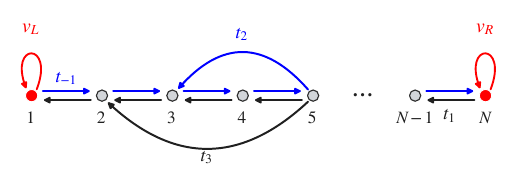}
\caption{Directed finite-range chain with $N$ sites. Arrows carry the
real hopping $t_n$ from site $j+n$ to $j$; arcs show representative
bulk hoppings, repeated along the chain. Red endpoints carry the
on-site controls $v_L,v_R$. Bulk hoppings are $t_{-1}=3/2$, $t_1=-9/44-\dd$,
$t_2=21/11$, and $t_3=-1$, where $\dd$ is the bulk detuning.
The omitted central segment is marked by dots.}
\label{fig:chain}
\end{figure}

\begin{equation}
\begin{aligned}
\hat{\mathcal H}_N^{v_L,v_R}&=
\sum_{j,n}
t_n\hat c_j^\dagger\hat c_{j+n}+v_L\hat n_1+v_R\hat n_N,\\[2pt]
H_\dd(\beta)&=t_{-1}\beta^{-1}+t_1\beta+t_2\beta^2+t_3\beta^3.
\end{aligned}
\label{eq:model}
\end{equation}

Here $\hat c_j^\dagger$ and $\hat c_j$ create and annihilate a particle
at site $j$; $\hat n_j=\hat c_j^\dagger\hat c_j$ counts bosons or fermions.
The sum uses signed displacements $n\in\mathcal S=\{-1,1,2,3\}$
and sites with $1\le j,j+n\le N$; real $t_n$ hops from
site $j+n$ to $j$. Unpaired longer-range hoppings make the chain non-Hermitian.
The bulk on-site term is $t_0=0$.
The real endpoint potentials $v_L,v_R$ are independent of $N$.
Energies, hoppings, and potentials are in units of $|t_3|=1$;
$\dd$ is the bulk detuning entering $t_1$.
The one-particle matrix is $H_N^{v_L,v_R}$;
$H_N=H_N^{0,0}$ is bare [Fig.~\ref{fig:chain}].
The Bloch factor $\beta=e^{ik}$ has complex wave number $k$ and amplitude
proportional to $\beta^j$.

The representative hoppings in Fig.~\ref{fig:chain} belong to an analytic
cooperative family with $r=t_{-1}$ and $1<r<1+2/\sqrt3$.
The remaining hoppings place a quadratic modulus contact at
$\dd=0$ and critical energy $\Ec$; their construction is given in
Supplemental Material (SM), Sec.~S1 \cite{Supplemental}.
Small real translation-invariant hopping changes preserve cooperation
after relocating the critical point $(\dd_c,\Ec)$
\cite{Supplemental}.

\begin{figure}[t]
\includegraphics[width=\columnwidth]{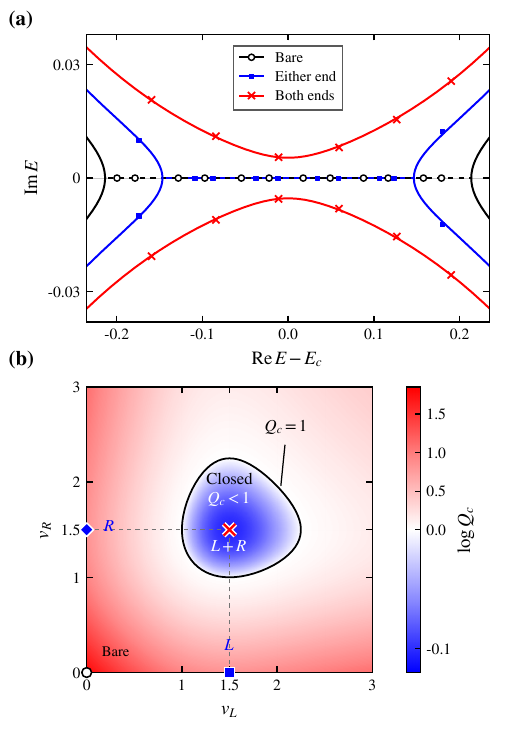}
\caption{Cooperative closure at fixed bulk coefficients.
(a) Local spectra at $N=256$, $\dd=0$, and $E_c=71/44$.
With $r=3/2$, black circles denote bare $(v_L,v_R)=(0,0)$,
blue squares either end $(r,0)$ or $(0,r)$, and red crosses both ends $(r,r)$.
Their displayed-window real-level counts are $N_{\mathbb R}=12,8,0$, respectively.
Solid curves are asymptotic complex branches derived in the Supplemental
Material; alternating black and blue dashes distinguish overlapping
real-axis allowed segments.
(b) $\log Q_c$ versus $v_L,v_R$, where $Q_c$ is the critical
conjugate-pair amplitude relative to the real term.
The black $Q_c=1$ contour bounds the asymptotic closing region $Q_c<1$.}
\label{fig:boundary}
\end{figure}

Figure~\ref{fig:boundary}(a) shows the local spectral change with bulk
hoppings fixed. Bare and either optimal single-end boundaries retain real
levels near $\Ec$; applying both potentials moves this local group into
complex-conjugate branches. The identical single-end spectra follow from
the endpoint symmetry of the characteristic determinant.
Figure~\ref{fig:boundary}(b) shows an open closing region, including
unequal endpoint potentials. Boundary matching explains why a second
controlled end is necessary.

\suppressfloats[t]

\emph{Boundary matching.---}
The spatial waves solve $H_\dd(\beta)=E$, where $E$ is energy.
At the contact, the real root $\beta_0=-1$ and a conjugate pair
$\beta_\pm$ have equal modulus; a fourth real root is separated.
All four roots are distinct \cite{Supplemental}.
Equal-modulus junctions occur in spectral graphs
\cite{XiongHu2024,BottcherEtAl2021}.
Here the logarithmic modulus gap $\Delta=\log|\beta_0/\beta_+|$
has the leading local form $\Delta\simeq-A\dd-B(E-\Ec)^2$.
Positive $A$ and $B$ measure detuning response and contact curvature.
The spatial attenuation rates meet without a spatial-root collision.

Matching the spatial waves at the endpoints fixes the finite-chain
quantization condition. Real energies are zeros of
$\cD_N(E)=\det(EI-H_N^{v_L,v_R})$, where $I$ is the identity matrix.
For nonsingular endpoints $v_L,v_R\ne-r$, division by the real
contribution gives
\begin{equation}
F_N=1+Qe^{N\Delta}\cos\Theta_N+R_N.
\label{eq:mechanism}
\end{equation}
Here $F_N$ is the determinant normalized by its real contribution,
$Q=Q(E,\dd)>0$ is the relative amplitude of the conjugate pair,
$Q_c=Q(\Ec,0)$ its critical value, and
$\Theta_N=N\phi+\psi$ is its phase; $\phi$ and $\psi$ are the phase
per site and boundary phase shift.
The remainder $R_N$ is exponentially small \cite{Supplemental}.
A real level requires the oscillatory term to cancel the real term:
its envelope must reach one at leading order.
Endpoint potentials tune this envelope through the boundary matching
amplitudes while keeping the bulk propagation factors fixed.

\emph{Why two ends are needed.---}
The matching factors from the two endpoints multiply.
For a real potential, the smallest individual factor is nonzero:
one endpoint has a finite amplitude-reduction capacity.
Let $Q_0$ be the bare critical amplitude ratio, and let
$\sigma\in(0,1)$ be the best single-end reduction factor, fixed by
the relative bulk phase.
Optimizing fixed nonsingular real potentials gives the smallest
attainable critical ratio
\begin{equation}
Q_{c,\min}^{(m)}=Q_0\sigma^m,\qquad m=0,1,2.
\label{eq:class}
\end{equation}
Here $m$ counts independently controlled ends; the others are bare.
SM Sec.~S2 states the matching factors and applicable scalar-chain assumptions.
At $\dd=0$, the strict interval $\sigma^{-1}<Q_0<\sigma^{-2}$
therefore identifies the cooperative class: one end cannot bring the
envelope below one, while two ends can.
Increasing a single-end potential cannot bypass this lower bound;
the second end supplies a further multiplicative reduction without
a direct link between the endpoints.

This strict condition holds throughout the analytic family
\cite{Supplemental}. Each controlled end is optimal at $v=+r$;
the excluded opposite value $v=-r$ cancels the real contribution.
Every fixed nonsingular real single-end potential preserves the population;
two ends close it in an open region, without requiring equal potentials
[Fig.~\ref{fig:boundary}(b)].

\begin{figure}[t]
\includegraphics[width=\columnwidth]{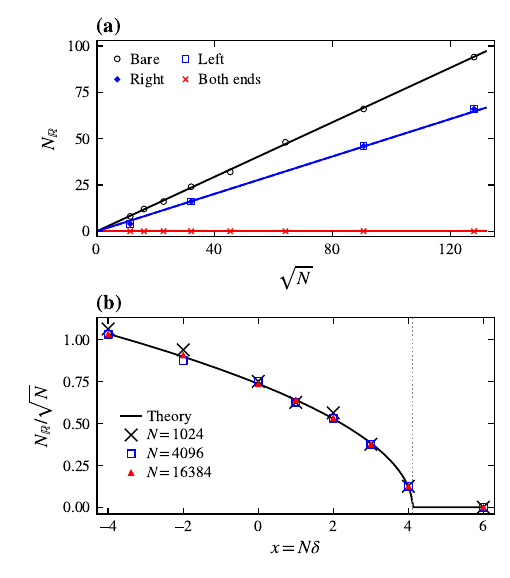}
\caption{A growing population with vanishing spectral fraction.
$\NR$ counts local real eigenvalues and $N$ is chain length.
Symbols give real-root counts from the exact finite determinant; curves
use analytic coefficients. Size-dependent numerical windows enclose
the predicted population (Supplemental Material).
(a) Bare, optimal left-, right-, and two-end boundaries at $\dd=0$,
compared with Eq.~\eqref{eq:count}.
Nested blue squares and diamonds distinguish coincident single-end counts.
(b) Bare-boundary crossover versus $x=N\dd$.
Black crosses, blue squares, and red triangles denote
$N=1024,4096,16384$, respectively.
Nested markers distinguish points that overlap in the scaled coordinates
$(x,\NR/\sqrt N)$ without displacing them.
The solid line is the leading scaled count from Eq.~\eqref{eq:count};
the vertical dashed line is the closing threshold $\log Q_0/A$,
where $Q_0$ is the bare ratio and $A$ the detuning coefficient.}
\label{fig:counts}
\end{figure}

\emph{How many modes are controlled.---}
The quadratic contact confines real levels to an energy window of width
$N^{-1/2}$. Their phase accumulates along the chain, giving level spacing
of order $N^{-1}$. The levels become dense faster than the window shrinks,
so the window contains a growing $\sqrt N$ population.
Let $\NR$ count real eigenvalues, with algebraic multiplicity, in a
fixed small interval around $\Ec$. Define $b=|\partial_E\phi|_c>0$
as the critical phase slope and $x=N\dd$ as scaled bulk detuning;
the subscript $c$ denotes the contact.
For fixed $\log Q_c-Ax>0$, the count is
\begin{equation}
\NR=\frac{2b}{\pi\sqrt B}\sqrt{\log Q_c-Ax}\,\sqrt N+O(1).
\label{eq:count}
\end{equation}
For fixed $\log Q_c-Ax<0$, the local population is eventually zero.
At $\log Q_c-Ax=0$, the count is $O(1)$ and depends on the phase sequence;
threshold equality does not guarantee closure.
These asymptotic results require fixed nonsingular boundaries
\cite{Supplemental}.
Figure~\ref{fig:counts} tests the predicted size dependence and detuning
crossover. Thus $\NR/N$ vanishes despite the growing count.
Closure is local; other real or isolated boundary levels are not excluded.

\suppressfloats[t]

\begin{figure*}[t]
\includegraphics[width=\textwidth]{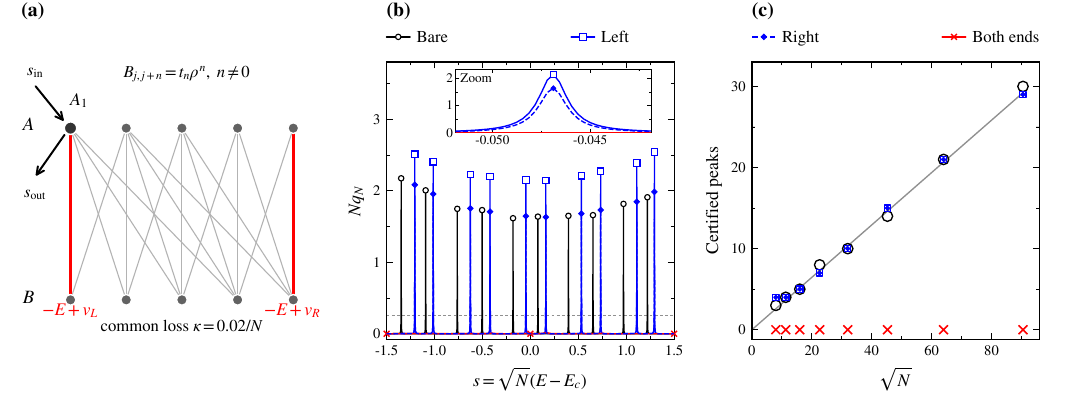}
\caption{Fixed-port passive reflection of the cooperative switch.
(a) Passive Hermitian double chain with port $A_1$ and endpoint interchain controls.
(b) Scaled amplitude defect $Nq_N$ versus
$s=\sqrt N(E-\Ec)$ at $N=1024$.
All settings share $\kappa=0.02/N$, port, and blind grid;
the gray dashed line marks the analytic separation level $Nq_N=0.25$.
Inset: coincident left- and right-end resonances with unequal port weights,
on the same $(s,Nq_N)$ axes and scan points.
(c) Certified resonance counts on sampled profiles through $N=8192$ in
$|s|<1.5$ match independent root counts; line: leading phase-cell count.
The profile certificate uses scaled support/seed thresholds 0.20/0.40
and preserves counts under pointwise amplitude error $0.09/N$.
$E$ tunes couplings at fixed probe carrier.}
\label{fig:readout}
\end{figure*}

\emph{A stable local consequence.---}
A real level need not be visible at a fixed port, and absence of real
levels does not bound the response background.
Skin modes have been observed in light and quantum-walk systems
\cite{WeidemannEtAl2020,XiaoEtAl2020}, and in electrical or mechanical networks
\cite{HelbigEtAl2020,GhatakEtAl2020,BrandenbourgerEtAl2019}.
Skin accumulation also affects damping \cite{SongYaoWang2019Damping}.
Hermitian devices can support effective non-Hermitian dynamics
\cite{WangClerk2019,McDonaldEtAl2018}.
For this chain, uniform damping with a fixed positive passivity margin makes
the joint endpoint contribution to a local response exponentially small
\cite{Supplemental}.
We use an independent passive Hermitian double chain for the representative
model with $r=3/2$, $\dd=0$, and the four settings
$(v_L,v_R)=(0,0),(r,0),(0,r),(r,r)$.

The network has oscillator sites $A_j,B_j$.
Let $\rho(E)=-\beta_0(E)>0$, where $\beta_0(E)$ is the real spatial
root through $-1$, and directly implement
$H_{N,\rho}[j,j+n]=t_n\rho^n$ with the same endpoint potentials.
Write $\mathsf B_N(E)=H_{N,\rho}^{v_L,v_R}-EI$.
Here $E$ is a hardware scan parameter: it tunes the rung coupling
$-E$ and the common bulk factors $\rho(E)$, at fixed probe carrier.
The real symmetric coupling matrix and its carrier-frequency
reflection defect, obtained from input-output theory
\cite{GardinerCollett1985,FanEtAl2003}, are
\begin{equation}
\begin{aligned}
\mathbb K_N(E)&=\begin{pmatrix}0&\mathsf B_N\\\mathsf B_N^T&0\end{pmatrix},\\
q_N(E)&=1-\varrho_N(E)
=\kappa^2\big[(\kappa^2I+\mathsf B_N\mathsf B_N^T)^{-1}\big]_{11}.
\end{aligned}
\label{eq:readout}
\end{equation}
Here $\mathbb K_N$ couples $2N$ modes, $\varrho_N$ is reflection amplitude,
and $q_N$ its real amplitude defect, not power.
The subscript 11 selects port $A_1$; $T$ denotes transpose.
All sites have total amplitude loss $\kappa=0.02/N$; at $A_1$ the
external port loss is $\kappa/2$. The internal loss allocation is given
in SM Sec.~S4. A fixed carrier above $\|\mathbb K_N\|$ makes all
oscillator frequencies positive. Every mode decays at rate $\kappa$,
without transient norm amplification.

The four settings share port, loss, and calibration.
In $|\sqrt N(E-\Ec)|<1.5$ at $\dd=0$, bounded finite-range couplings
are implemented directly.
The identity $\det\mathsf B_N=(-1)^N\cD_N$ maps real levels to
zero-energy crossings of this network. Port visibility requires more.
At each real root of the bare or either optimal single-end setting,
the $A_1$ weight is at least $c/N$.
For the optimal two-end setting, the smallest singular value of
$\mathsf B_N$ is at least $c'/N$ throughout the window.
Here $c,c'>0$ are independent of $N$. Together with port-resolved
inverse estimates, these bounds establish a strict peak-background
separation for sufficiently large $N$ \cite{Supplemental}.
Bare and optimal single-end settings support a $\sqrt N$ population
of resolvable resonance groups; the optimal two-end setting stays below
the separation level throughout the window.
Absolute amplitude contrast and stability margin both scale as $N^{-1}$.

Figure~\ref{fig:readout} compares a blind $0.01/N$ scan with independent
determinant roots in the narrower readout window.
Its finite sampled-profile counts use support and seed thresholds
$Nq_N=0.20,0.40$, distinct from the analytic separation level
$Nq_N=0.25$. A profile certificate preserves these counts under
amplitude error $0.09/N$ per point \cite{Supplemental}.
These finite-grid certificates do not supply an onset size for the
asymptotic bounds.

\emph{Discussion.---}
Boundary controls change the local real-mode count without changing the
local limiting bands. The minimum number of controlled ends is set by
boundary matching amplitudes, which these bands omit.
Vanishing normalized weight also hides the growing population from
thermodynamic spectral density.
Real-spectrum transitions and symmetry criteria are established in theory
\cite{BenderBoettcher1998,Mostafazadeh2002,ZhangEtAl2023Lieb} and photonic or circuit experiments
\cite{RuterEtAl2010,RegensburgerEtAl2012,SchindlerEtAl2011}.
Here, real eigenvalue pairs disappear through finite-matrix exceptional
points; the bulk spatial roots remain distinct.

The detuning crossover and the closed pairs' imaginary energies scale as
$N^{-1}$. An energy resolution coarser than this scale can count a
growing near-real population even after strict real levels have vanished;
SM Sec.~S5 gives the resolution threshold \cite{Supplemental}.

The shrinking $N^{-1}$ signal and stability margin impose readout costs.
Sufficient auxiliary budgets are $O(N^{-2})$ Hermitian operator-norm calibration,
$O(N\log N)$ settling per point, and $O(N^{7/2}\log N)$ photons
under ideal coherent homodyne with vacuum reservoirs, for a conservative
$O(N^{-2})$ scan at fixed overall error probability
\cite{Supplemental,ClerkEtAl2010}. Loss and port errors need separate bounds; thermal noise,
inefficiency, and technical noise are outside this ideal budget.
Stabilization and sensing advantage in the original device remain open.
Spectral sensitivity \cite{Wiersig2014,BudichBergholtz2020}
alone does not establish precision; noise and resource analysis is needed
\cite{Langbein2018,ZhangEtAl2019QuantumNoise,ChenJinLiu2019,LauClerk2018Sensing,McDonaldClerk2020}.

Small real translation-invariant hopping changes preserve the cooperative
class after relocating the bulk critical point. Threshold equality and
coefficient-zero endpoints need separate treatment \cite{Supplemental}.
For the representative chain's four settings, real on-site errors
$\xi_j$ with a sufficiently small $N$-independent
budget $\sum_j|\xi_j|\le\tau$, $\tau>0$, preserve closure and
bare and optimal single-end square-root counts in
$|E-\Ec|<(5/2)/\sqrt N$ at $\dd=0$ \cite{Supplemental}.
If every site is perturbed, worst-case $O(N^{-1})$ per-site amplitude
suffices. Fixed-strength extensive disorder, general complex hoppings,
and arbitrary multiband systems remain open.
Localization with asymmetric hopping
\cite{HatanoNelson1996,WangLiu2022UniformField}, localization in dissipative
surface states \cite{DingZhuangLiu2026Localization}, interacting bosonic phases
and doublons \cite{LiWuLiu2024Mott,WangLiu2026Doublons}, and Majorana statistics
\cite{ZhaoEtAl2021Majorana} involve additional physics beyond this scalar-chain criterion.

In summary, two separated real endpoint potentials can remove a local
$\sqrt N$ population of real modes that every fixed nonsingular real
single-end potential preserves in an open critical chain family.
The product of two bounded matching-amplitude reductions permits closure
in an open two-end region at fixed bulk. The controlled population grows
although its normalized weight vanishes; identical local limiting bands
need not imply identical control of finite-chain real spectra.
For a representative chain's four zero-detuning settings, an independent
passive Hermitian double chain gives a resolvable resonance population
at the same fixed local port and suppresses it under two-end control,
using a common coupling scan at fixed carrier.
Its shrinking $N^{-1}$ signal and stability margin impose explicit
resource costs. This local consequence does not establish stabilization
or sensing advantage in the original chain. The result identifies a
finite-chain control requirement absent from thermodynamic spectral
density: how many endpoints are needed to remove a growing real-mode population.

\begin{acknowledgments}
This work is supported by the Zhiyuan Science Foundation of BIPT (No. 2026210).

We thank Tianyu Xie of Xiamen University for providing computing resources.
\end{acknowledgments}

\bibliographystyle{apsrev4-2}
\raggedright
\nocite{BeenakkerEtAl2013Parity,BogoyaEtAl2025,EdelmanEtAl1994,FeinbergZee1997HermitianReduction,HegdeVishveshwara2016,LauClerk2018Sensing,PekertenEtAl2019,ShapiroStampach2017,ShapiroStampachCorrection2023,YiYang2025}
\bibliography{refs}
\onecolumngrid
\clearpage
\justifying
\setcounter{equation}{0}
\setcounter{figure}{0}
\setcounter{table}{0}
\setcounter{section}{0}
\setcounter{subsection}{0}
\renewcommand{\theequation}{S\arabic{equation}}
\renewcommand{\thefigure}{S\arabic{figure}}
\renewcommand{\thetable}{S\Roman{table}}
\renewcommand{\thesection}{S\arabic{section}}
\renewcommand{\theHequation}{supp.\arabic{equation}}
\renewcommand{\theHfigure}{supp.\arabic{figure}}
\renewcommand{\theHtable}{supp.\arabic{table}}
\renewcommand{\theHsection}{supp.\arabic{section}}
\setcounter{secnumdepth}{2}
\begin{center}
{\bfseries Supplemental Material for}\par\medskip
{\bfseries Cooperative Boundary Switching of Mesoscopic Real Spectra in Non-Hermitian Chains}\par\medskip
Chengxi Li$^{1,2}$, Hao Zhu$^{1,2}$, Tie-Fu Zhang$^{2,3}$, Wanzi Sun$^{2,4}$\par\medskip
{\small $^{1}$Beijing Institute of Petrochemical Technology, Beijing 102617, China}\par
{\small $^{2}$Beijing National Laboratory for Condensed Matter Physics, Institute of Physics, Chinese Academy of Sciences, Beijing 100190, China}\par
{\small $^{3}$Faculty of Science, Kunming University of Science and Technology, Kunming 650500, China}\par
{\small $^{4}$Jining University, Qufu, Shandong 273155, China}\par
\smallskip
{\small Corresponding author: \href{mailto:haozhu4012@163.com}{haozhu4012@163.com}}
\end{center}
The derivations below follow the Letter from the boundary-dependent
determinant to the minimum endpoint requirement, real-level count, and
passive local response. The last sections give the complex branches,
resolution and spatial-error bounds, and the numerical methods used
in Figs.~2--4 of the Letter.

\section{Finite-chain determinant and quadratic modulus contact}
\label{sec:exact}
\label{sec:uniform}

The distinction in Eqs.~(1) and (2) of the Letter is between bulk
exponential factors and their boundary-dependent coefficients. In the
one-particle basis, the open-chain matrix obeys
$H_{j,j+n}=t_n$ for $n=-1,1,2,3$, with $1\le j,j+n\le N$.
The lattice spacing and $|t_3|$ are one. We use the family
\begin{align}
H_\dd^{(r)}(\beta)&=r\beta^{-1}
+[(r-1)(1-2a)-\dd]\beta+(r-1+2a)\beta^2-\beta^3,\nonumber\\
a(r)&=\frac{3r^2+1}{6r+2},\qquad
1<r<1+\frac2{\sqrt3},\qquad \Ec(r)=1+r(2a-1).
\label{eq:Sfamily}
\end{align}
Here $\beta^j$ is the spatial amplitude and $\dd$ is bulk detuning.
Real endpoint potentials are added as
$H_N^{v_L,v_R}=H_N+v_L|1\rangle\langle1|+v_R|N\rangle\langle N|$.
Unless specified otherwise, numerical constants refer to $r=3/2$,
for which $(t_1,t_2,t_3)=(-9/44-\dd,21/11,-1)$ and $\Ec=71/44$.

The spatial roots solve the inverse-band polynomial
\begin{equation}
P(\beta;E,\dd)=\beta[H_\dd^{(r)}(\beta)-E]
=r+t_1\beta^2+t_2\beta^3-\beta^4-E\beta.
\label{eq:Spoly}
\end{equation}
At $\dd=0$ it has the exact critical factorization
\begin{equation}
P(\beta;\Ec,0)=-(\beta+1)(\beta-r)(\beta^2-2a\beta+1).
\label{eq:Sfactor}
\end{equation}
The roots are $-1,e^{\pm i\arccos a},r$, all simple in the stated
open family. Label their analytic continuations by
$\beta_0,\beta_\pm,\beta_3$; for real $E,\dd$, the first and last
are real and $\beta_-=\overline{\beta_+}$. The representative
conjugate roots are $(31\pm5i\sqrt{39})/44$.
The three active roots meet in modulus, not in value.

Write $D_N=\det(EI-H_N)$ and
$\cD_N=\det(EI-H_N^{v_L,v_R})$, where $I$ is the identity.
The banded Toeplitz determinant expansion
\cite{BottcherGrudsky2005,DuitsKuijlaars2008,KoekenbierSchulzBaldes2024}
gives
\begin{equation}
D_N=\sum_{j\in\{0,+,-,3\}}c_j\alpha_j^N,\qquad
c_j=\frac{\prod_{k\ne j}\beta_k}{\prod_{k\ne j}(\beta_k-\beta_j)},
\qquad \alpha_j=\frac r{\beta_j}.
\label{eq:Swidom}
\end{equation}
The root product is $-r$ and $D_0=1$; these fix the normalization
as the monic energy determinant. Deleting either endpoint leaves
a shorter truncation of the same bulk matrix. Multilinearity
therefore yields, for $N\ge2$,
\begin{equation}
\cD_N=D_N-(v_L+v_R)D_{N-1}+v_Lv_RD_{N-2}
=\sum_j\widetilde c_j\alpha_j^N,\qquad
\widetilde c_j=c_j
\left(1-\frac{v_L}{\alpha_j}\right)
\left(1-\frac{v_R}{\alpha_j}\right).
\label{eq:Sboundary}
\end{equation}
Endpoint control thus changes the relative amplitudes while
preserving the bulk bases. Where the active coefficients stay
nonzero, the local equal-modulus limiting curves are unchanged.
Isolated boundary levels elsewhere are outside this local statement.

For $v_L,v_R\ne-r$, the real coefficient $\widetilde c_0$ is
nonzero near the contact. A real potential also cannot cancel the
nonreal $\beta_+$ factor. Dividing by $\widetilde c_0\alpha_0^N$
combines the conjugate pair into a cosine:
\begin{align}
F_N&=1+e^{\ell_N}\cos\Theta_N+R_N,\qquad
\ell_N=\log Q+N\Delta,\nonumber\\
Q&=2|\widetilde c_+/\widetilde c_0|,\qquad
\Delta=\log|\beta_0/\beta_+|,\qquad
\Theta_N=N\phi+\psi,\nonumber\\
\phi&=\arg(\alpha_+/\alpha_0),\qquad
\psi=\arg(\widetilde c_+/\widetilde c_0),\qquad
R_N=\frac{\widetilde c_3}{\widetilde c_0}
\left(\frac{\alpha_3}{\alpha_0}\right)^N .
\label{eq:Snorm}
\end{align}
The phases use continuous local branches. A real zero requires
$e^{\ell_N}\ge1+R_N$: the oscillatory term must be able to cancel
the real term. This amplitude condition carries information absent
from the limiting curves.

\begin{figure}[t]
\includegraphics[width=.76\textwidth]{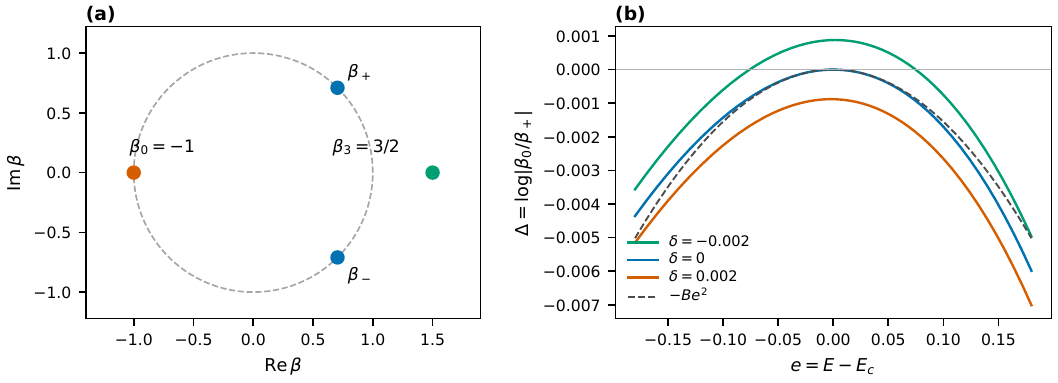}
\caption{Root geometry and quadratic modulus gap of the representative
chain. (a) The roots at $\dd=0$, $E=\Ec=71/44$: three have unit modulus
and the fourth is $3/2$. (b) $\Delta=\log|\beta_0/\beta_+|$ from analytic
roots for the indicated detunings, compared with $-B(E-\Ec)^2$ at
$\dd=0$. No spatial roots coalesce.}
\label{fig:Sroots}
\end{figure}

Implicit differentiation gives
$\beta_{j,E}=\beta_j/P_\beta(\beta_j)$ and
$\beta_{j,\dd}=\beta_j^2/P_\beta(\beta_j)$.
For $e=E-\Ec$, the modulus gap and phase rate are
\begin{align}
\Delta&=-A\dd-Be^2+O(\dd^2,e\dd,e^3),\nonumber\\
A&=\frac{11}{25},\qquad B=\frac{484}{3125},\qquad
b=|\phi_{E,c}|=\frac{88\sqrt{39}}{1625}.
\label{eq:Sgap}
\end{align}
Here $A$ is the detuning response, $B>0$ the contact curvature,
$b>0$ the phase slope, and the subscript $c$ denotes the contact.
Figure~\ref{fig:Sroots} illustrates the distinct roots and quadratic
gap; this geometry differs from an inverse-root collision or the
onset of a wholly real non-Bloch band \cite{HuEtAl2024,ChenZhang2023}.

The fourth base has relative modulus $2/3$ at the representative
contact, and $1/r<1$ throughout the family. Root separation and
nonzero coefficients give a fixed closed local neighborhood with
$|\alpha_3/\alpha_0|\le\rho_{\rm sub}<1$ and bounded derivatives.
Consequently,
\begin{equation}
|\partial_E^jR_N|\le C_jN^j\rho_{\rm sub}^N,\qquad j=0,1,2.
\label{eq:Srem}
\end{equation}
The constants are independent of $N$ and uniform on compact endpoint
sets avoiding $v_i=-r$ and compact subintervals of the open family.
The moduli and phases in Eq.~\eqref{eq:Snorm} define real analytic
functions on the real axis. Complex-energy calculations use their
local holomorphic extensions.

\section{Minimum endpoint control and the open chain family}
\label{sec:class}
\label{sec:cooperation}

Equation~(3) of the Letter follows by optimizing the coefficient
factor in Eq.~\eqref{eq:Sboundary}. More generally, consider a real
scalar Toeplitz chain with one real dominant base $\alpha_A\ne0$
and a conjugate pair $\alpha_B,\overline{\alpha_B}$ of equal
modulus at the contact. Their coefficients $c_A,c_B$ are nonzero.
The three-term analytic contact, subdominant gap, and nonzero phase
slope specified below are required, together with the endpoint
identity in Eq.~\eqref{eq:Sboundary}.

Set $Q_0=2|c_B/c_A|$, $\theta=\arg(\alpha_B/\alpha_A)$, and
$u_i=v_i/\alpha_A$, all at the contact. The endpoint identity gives
\begin{align}
Q_c&=Q_0p(u_L)p(u_R),\qquad
p(u)=\frac{|1-ue^{-i\theta}|}{|1-u|},\quad u\ne1,\nonumber\\
p(u)^2&=\sigma^2+\frac{1-\cos\theta}{2}
\left(\frac{u+1}{u-1}\right)^2,\qquad
\sigma=|\cos(\theta/2)|\in(0,1).
\label{eq:SclassQ}
\end{align}
The nonnegative square proves the unique minimum $p(-1)=\sigma$.
The best potential is $v=-\alpha_A$; the excluded opposite value
$v=\alpha_A$ cancels the real coefficient. For $m=0,1,2$ controlled
ends, with the others bare, the minimum is $Q_0\sigma^m$.
At fixed $x=N\dd$, the envelope can be lowered strictly below one
when $Q_0\sigma^m<e^{Ax}$. At $x=0$ the cooperative class is
\begin{equation}
Q_0\sigma>1,\quad Q_0\sigma^2<1
\quad\Longleftrightarrow\quad \sigma^{-1}<Q_0<\sigma^{-2}.
\label{eq:Sclasscriterion}
\end{equation}
One endpoint has a finite reduction capacity; the product of two
independent reductions can cross the cancellation threshold.

For the family in Eq.~\eqref{eq:Sfamily}, let
$d(r)=-3r^2+6r+1>0$. Differentiating its factored inverse polynomial
gives
\begin{align}
A(r)&=\frac{3r+1}{2(r+1)^2},\qquad
B(r)=\frac{r(3r+1)^2(9r-1)}{6(r+1)^7},\nonumber\\
b(r)^2&=\frac{4r^2(3r+1)^2}{3(r+1)^6d(r)}.
\label{eq:Sfamilyconstants}
\end{align}
These expressions follow from the implicit-root derivatives above:
reducing $1/P_\beta$ and
$(P_\beta-\beta P_{\beta\beta})/P_\beta^3$ modulo
$\beta^2-2a\beta+1$, then taking the real part at $\beta_\pm$,
gives $\Delta_E=0$ and $\Delta_{EE}=-2B$.
All three constants are positive. The coefficient ratio and phase
factor are
\begin{align}
Q_0(r)&=\frac{2(3r+1)}{\sqrt{d(r)}},\qquad
\sigma^2=\frac{d(r)}{4(3r+1)},\nonumber\\
Q_{c,\min}^{(1)}&=\sqrt{3r+1}>1,\qquad
Q_{c,\min}^{(2)}=\frac{\sqrt{d(r)}}2<1.
\label{eq:Sfamilymin}
\end{align}
Here $\alpha_A=-r$, so every controlled end is optimal at $+r$.
The last inequality uses $d(r)=4-3(r-1)^2<4$.
Every fixed nonsingular real single-end potential therefore leaves
the local population open, while an open two-end region closes it.
The estimates are uniform on compact family subintervals. At $r=1$
the fourth base also becomes equimodular; at $r=1+2/\sqrt3$ the
conjugate roots collide. Neither endpoint is covered.

For $r=3/2$, Eq.~\eqref{eq:SclassQ} becomes the phase diagram
in Fig.~2(b) of the Letter:
\begin{equation}
Q_c(v_L,v_R)=\frac{22}{\sqrt{13}}
\sqrt{f(v_L/r)f(v_R/r)},\qquad
f(u)=\frac{1-2(31/44)u+u^2}{(1+u)^2}.
\label{eq:SQ}
\end{equation}
The single-end and double-end minima are $\sqrt{11/2}$ and
$\sqrt{13}/4$. The strict region $Q_c<1$ includes
$(v_L,v_R)=(1.425,1.575)$, where $Q_c=0.9046441012$.
Thus closure needs neither equal potentials nor a direct link
between endpoints. Equality is treated separately in the next section.

\subsection{Changing the bulk hoppings}
\label{sec:perturb}

For small real translation-invariant changes within a fixed hopping
range, simple roots, nonzero coefficients and the subdominant gap
persist. The contact must move with the hoppings. The equations
$\Delta=0$, $\Delta_E=0$ have Jacobian
\begin{equation}
\frac{\partial(\Delta,\Delta_E)}{\partial(\dd,E)}
=\begin{pmatrix}-A&0\\ \Delta_{E\dd,c}&-2B\end{pmatrix},
\qquad \det=2AB\ne0 .
\end{equation}
The implicit-function theorem supplies a nearby contact with
$A,B,b>0$, where the strict endpoint inequalities still hold.
The old energy and detuning are not retained under this continuation.
Complex hoppings, extensive disorder and arbitrary multiband
matrices need separate analyses.

\section{Mesoscopic counting and the two boundary exceptions}
\label{sec:count}
\label{thm:regular}

The window width and phase spacing give the physical origin of
Eq.~(4) of the Letter. To control the count as well as its scaling,
assume the analytic three-term form in Eq.~\eqref{eq:Snorm},
nonzero real and conjugate coefficients, the derivative remainder
bounds in Eq.~\eqref{eq:Srem}, and
\begin{equation}
\Delta_c=\Delta_{E,c}=0,\quad
\Delta_{\dd,c}=-A<0,\quad \Delta_{EE,c}=-2B<0,\quad
|\phi_{E,c}|=b>0.
\label{eq:Shyp}
\end{equation}
The family above satisfies these conditions. Count roots with
algebraic multiplicity in a sufficiently small fixed interval
$|E-\Ec|<\epsilon$. At fixed $x=N\dd$, put
$K=\log Q_c-Ax$. Expansion gives
\begin{align}
\ell_N&=K-BNe^2+O(|e|+N|e|^3+N^{-1}),\nonumber\\
\ell_N''&=-2BN+O(N|e|+1),\qquad
|\Theta_N'|\ge bN/2,\qquad \Theta_N''=O(N).
\label{eq:Selltaylor}
\end{align}
A zero requires $e^{\ell_N}\ge1+R_N$, confining it to
$e=O(N^{-1/2})$. For $K>0$, the two envelope edges are
\begin{equation}
e_{\pm,N}=\pm\sqrt{K/(BN)}+O(N^{-1}).
\label{eq:Sedges}
\end{equation}
For $K<0$, the maximum $\ell_N=K+O(N^{-1})$ stays below zero
and there are no local real roots for sufficiently large $N$.

For $K>0$ only negative cosine half-waves can cancel the real term.
On such a half-wave define
$G_N=\ell_N+\log(-\cos\Theta_N)-\log(1+R_N)$.
Its curvature is
\begin{equation}
G_N''=\ell_N''-(\Theta_N')^2\sec^2\Theta_N
-\Theta_N''\tan\Theta_N-\partial_E^2\log(1+R_N).
\label{eq:Sconcavity}
\end{equation}
Since $|\tan\theta|\le\frac12\sec^2\theta$, the negative
$N^2\sec^2\Theta_N$ term dominates the phase curvature, envelope
curvature and remainder. Thus $G_N$ is strictly concave for large $N$
and tends to $-\infty$ at both ends. A complete allowed half-wave
has exactly two roots, a forbidden one has none, and a tangent one
has a double root.

At a half-wave center, $\Theta_N=(2k+1)\pi$,
$G_N=\ell_N+O(\rho_{\rm sub}^N)$ and $G_N'=O(\sqrt N)$.
The maximum shifts by $O(N^{-3/2})$ and its value by $O(N^{-1})$.
Near an envelope edge, center spacing is $O(N^{-1})$ and the
envelope slope is of order $\sqrt N$. Only a bounded number of
edge half-waves change their center-based classification.
Across the allowed window the phase advance is
$2b\sqrt{K/B}\sqrt N+O(1)$. There is one negative half-wave
per $2\pi$ and two roots per allowed half-wave, so
\begin{equation}
\NR=\frac{2b}{\pi\sqrt B}\sqrt K\sqrt N+O(1),\qquad K>0.
\label{eq:Scount}
\end{equation}
The error includes the edge cells and algebraic multiplicity at
tangencies. This is a fixed-interval local count, not a whole-band
count. It is uniform on compact parameter sets separated from
$K=0$, coefficient zeros and the family endpoints.

\subsection{Threshold equality}
\label{sec:equality}

At $K=0$ the cancellation window narrows further. Define
$t=Ne$, $q_E=(\log Q)_{E,c}$,
$q_\dd=(\log Q)_{\dd,c}$,
$d_{11}=\Delta_{E\dd,c}$ and
$d_{02}=\Delta_{\dd\dd,c}/2$. On compact $t$ sets,
\begin{equation}
\ell_N=\frac{h_x(t)}N+O(N^{-2}),\qquad
h_x(t)=-Bt^2+(q_E+d_{11}x)t+q_\dd x+d_{02}x^2 .
\label{eq:Sthreshold}
\end{equation}
Completing the square restricts the possible roots to $e=O(N^{-1})$,
which contains only $O(1)$ half-waves. Their centers satisfy
\begin{equation}
t_{k,N}=\frac{(2k+1)\pi-N\phi_c-\phi_{\dd,c}x-\psi_c}{\phi_{E,c}}
+O(N^{-1}).
\label{eq:Scenters}
\end{equation}
If $h_x(t_{k,N})$ at a half-wave center is bounded away from zero,
a positive value yields a real pair and a negative value yields no real pair.
Centers at zero to this order need higher Taylor terms.
The count remains bounded but depends on the discrete phase sequence;
threshold equality is not a universal closing condition.

\subsection{An exactly singular endpoint}
\label{sec:singular}
\label{thm:singular}

Set exactly $m=1$ or $2$ endpoints to $-r$ in the representative
chain, leaving any other endpoint fixed in a compact nonsingular set.
The factor $1+\beta_0$ has a simple zero at $e=\dd$, since
$H_\dd(-1)=\Ec+\dd$. With $\kappa_\beta=44/375$,
$1+\beta_0=-\kappa_\beta(e-\dd)+O((|e|+|\dd|)^2)$.
Normalize by the bare base $\alpha_0^N$, not by the zero coefficient.
Then
\begin{equation}
\widetilde c_0=C_A(e-\dd)^m[1+O(|e|+|\dd|)],\qquad
\widetilde c_+=C_B[1+O(|e|+|\dd|)],\qquad C_AC_B\ne0.
\label{eq:Ssingcoef}
\end{equation}
On either side of the coefficient zero, the relative-amplitude
logarithm is
\begin{equation}
L_N=\log C-Ax-BNe^2-m\log|e-\dd|+o(1),\qquad C=2|C_B/C_A|.
\label{eq:Ssingamp}
\end{equation}
For bounded fixed $x$, the outer edge $w_N$ therefore obeys
\begin{equation}
BNw_N^2=\frac m2\log N-\frac m2\log\log N+O(1).
\label{eq:Ssingwidth}
\end{equation}
The Taylor error is uniform there because $N|e|^3=o(1)$.

The central region $|e-\dd|\le L/N$, with sufficiently large
fixed $L$, contains only $O(1)$ roots. In the rescaled coordinate,
the oscillatory coefficient and phase speed stay nonzero; its
derivative is bounded away from zero near cosine zeros, and its
value is bounded away from zero outside those neighborhoods.
The real and subdominant terms vanish with their first derivatives.
Outside this region the real term has a fixed sign on either side.
The extra envelope curvature $m/(e-\dd)^2\le mN^2/L^2$ is
made smaller than the negative phase-square bound by choosing $L$.
The half-wave argument then applies with bounded edge corrections.
The phase length is $2bNw_N+O(\log N)$, giving
\begin{equation}
\NR=\frac{2b}{\pi}\sqrt{\frac m{2B}}\sqrt{N\log N}
\left[1+O\!\left(\frac{\log\log N}{\log N}\right)\right].
\label{eq:Ssingcount}
\end{equation}
This exception requires the potential to equal $-r$ exactly.
Approaching it with $N$ introduces another crossover.

\section{A fixed-port passive readout of the cooperative switch}
\label{sec:auxreadout}
\label{sec:readout}

The real levels counted above do not by themselves guarantee a local
signal. A mode can have negligible weight at the measured port, and
a nonzero determinant in the closed setting need not give a small
response background. This section derives both bounds for the
auxiliary network in Eq.~(5) and Fig.~4 of the Letter. Its construction
also addresses a separate obstacle in the original chain: at a fixed
positive passivity margin, the local response carrying information
about both distant endpoints is exponentially small.

\subsection{Joint endpoint response at a fixed passivity margin}
\label{sec:connectedresponse}

For the original linear system
$i\dot a=(H_N^{v_L,v_R}-i\gamma I)a+f e^{-i\omega t}$, define
$G^{v_L,v_R}(E)=(EI-H_N^{v_L,v_R})^{-1}$ at the common
$E=\omega+i\gamma$. Here $\omega$ is drive frequency and $\gamma$
uniform amplitude loss. At a site $i$ used as both source and detector,
the response that removes the effects of either endpoint alone is
\begin{equation}
\mathcal C_i=G_{ii}^{v_L,v_R}-G_{ii}^{v_L,0}
-G_{ii}^{0,v_R}+G_{ii}^{0,0}.
\label{eq:Sconnected}
\end{equation}
All four settings have the same bulk, drive frequency and loss.
Let $\gamma_{\rm pass}(N)=\lambda_{\max}[(H_N-H_N^\dagger)/(2i)]$.
Real endpoint potentials leave this passivity cost unchanged.
For $\gamma\ge\sup_N\gamma_{\rm pass}(N)+\epsilon_{\rm pass}$,
where $\epsilon_{\rm pass}>0$ is fixed, dissipativity gives
$\|G\|_2\le1/\epsilon_{\rm pass}$.

Choose a fixed $\chi>0$ so that, for either sign,
$\sum_n|t_n||e^{\pm\chi n}-1|\le\epsilon_{\rm pass}/2$.
Conjugation by $\operatorname{diag}(e^{\pm\chi j})$ changes the bulk
operator by at most this norm and leaves its diagonal potentials
unchanged. The Neumann-series bound for the conjugated resolvent
therefore yields
$|G_{ij}|\le(2/\epsilon_{\rm pass})e^{-\chi|i-j|}$.
Differentiating the resolvent with respect to the endpoint potentials
gives exactly
$\partial_{v_L}\partial_{v_R}G_{ii}
=G_{iN}G_{N1}G_{1i}+G_{i1}G_{1N}G_{Ni}$.
Each product travels a total distance $2(N-1)$.
Integration over the endpoint rectangle proves
\begin{equation}
|\mathcal C_i|\le
2|v_Lv_R|(2/\epsilon_{\rm pass})^3e^{-2\chi(N-1)}.
\label{eq:Sconnectedbound}
\end{equation}
The estimate is uniform in $i$, including sites that move with $N$,
and extends with a fixed prefactor to finitely many normalized local
ports. Its fixed passivity margin and fixed port normalization are
essential; a margin vanishing with $N$ or exponentially increasing
measurement resources require another estimate.

\subsection{A common finite-range network and a fixed carrier}

Hermitian doubling turns a real shifted matrix into a network with
signed singular values as its levels
\cite{FeinbergZee1997HermitianReduction,KawabataEtAl2019Symmetry,PorrasFernandezLorenzo2019}.
We use the representative chain at $\dd=0$ and exactly the four
settings $(v_L,v_R)=(0,0),(r,0),(0,r),(r,r)$, with $r=3/2$.
Continue the real spatial root from $\beta_0(\Ec)=-1$ and set
$\rho(E)=-\beta_0(E)>0$, $r_\rho(E)=r/\rho(E)$.
The directly implemented hopping matrix and its partner are
\begin{align}
[H_{N,\rho}]_{j,j+n}&=t_n\rho^n,\qquad
\mathsf B_N=H_{N,\rho}^{v_L,v_R}-EI,\nonumber\\
\mathbb K_N(E)&=
\begin{pmatrix}0&\mathsf B_N\\\mathsf B_N^T&0\end{pmatrix},
\qquad
I_N^{\rm aux}=\{|E-\Ec|<(3/2)/\sqrt N\}.
\label{eq:Sauxpartner}
\end{align}
The endpoint potentials remain additive diagonal entries of
$\mathsf B_N$. In the partner they are couplings between the
matching sites $A_j,B_j$ of its two sublattices.
The parameter $E$ tunes hardware couplings: every rung contains
$-E$, and the same bulk root calibrates all $t_n\rho(E)^n$
in each setting. It is independent of finite-chain states and
resonance positions. The probe stays at one common carrier.

The partner is real symmetric and obeys
$\Gamma\mathbb K_N\Gamma=-\mathbb K_N$, with
$\Gamma=\operatorname{diag}(I,-I)$.
In the order $A_1,B_1,A_2,B_2,\ldots$, it has half-bandwidth seven
and at most five links per site. On $I_N^{\rm aux}$,
$\rho=1+O(N^{-1/2})$; the couplings, $\rho'$ and
$\|\mathbb K_N'\|_2$ are bounded uniformly for large $N$.
A carrier $\Omega>\sup_{N,E}\|\mathbb K_N(E)\|_2$ makes
$\Omega I+\mathbb K_N$ a number-conserving oscillator network
with positive frequencies. Driving at $\Omega$ probes zero detuning
relative to $\mathbb K_N$. Energies and rates use the same hopping units.

The diagonal matrix
$D_\rho=\operatorname{diag}(1,\rho,\ldots,\rho^{N-1})$
satisfies $H_{N,\rho}^{v_L,v_R}=D_\rho^{-1}H_N^{v_L,v_R}D_\rho$.
Thus
\begin{equation}
\det\mathsf B_N=(-1)^N\cD_N,\qquad
\operatorname{spec}\mathbb K_N
=\{\pm\sigma_j(\mathsf B_N)\}_{j=1}^N.
\label{eq:Sauxspectralmap}
\end{equation}
Here $\sigma_j$ denotes a singular value.
The physical network uses the finite coefficients $t_n\rho^n$
directly; $D_\rho$ is only a device for proving the identity.
This spectrum is a singular-value embedding, and its local
measurable signal needs the additional estimates that follow.

\subsection{The closed-state background and the open-state port weight}

For a segment of $n$ sites, let $D_n,D_n^L,D_n^R$ denote the
bare, left-end and right-end energy determinants, with $D_0^X=1$.
For $n\ge1$, $D_n^L=D_n-v_LD_{n-1}$ and
$D_n^R=D_n-v_RD_{n-1}$.
Normalize them by $\alpha_0=-r_\rho$:
$\mathfrak d_n^X=D_n^X/\alpha_0^n$ and
$\mathfrak d_N^{LR}=\cD_N/\alpha_0^N$.
Their bases in Eq.~\eqref{eq:Swidom} are
$1,\beta_0/\beta_+,\beta_0/\beta_-,\beta_0/\beta_3$,
of modulus at most one near the real contact at $\dd=0$.

At $\Ec$, the absolute coefficient sums are
$1.60189958\ldots$ for bare segments and
$1.35351317\ldots$ for either optimal one-end segment.
The two-end leading coefficient is $88/125$, its relative
oscillatory amplitude is $\sqrt{13}/4$, and its fourth base has
modulus $2/3$. Continuity and this strict gap give, for some onset
size $N_0$, all $N\ge N_0$, $E\in I_N^{\rm aux}$ and $0\le n\le N$,
\begin{align}
|\mathfrak d_n|&\le M_0=17/10,\qquad
|\mathfrak d_n^{L,R}|\le M_1=29/20,\nonumber\\
\mathfrak d_N^{LR}&\ge d_*=3/50,\qquad
r_\rho\ge r_*=7/5.
\label{eq:Sauxminorbound}
\end{align}
The full-length lower bound uses the limiting margin
$(88/125)(1-\sqrt{13}/4)=0.0694229755\ldots>3/50$.
No numerical value of $N_0$ is certified by this continuity argument.

The upper Hessenberg recurrence supplies the exact inverse
\begin{align}
G_{ij}
&=r_\rho^{i-j}\left[
\frac{D_{j-1}^LD_{N-i}^R}{\cD_N}
-\mathbf1_{j>i}D_{j-i-1}\right]\nonumber\\
&=\frac{(-1)^{j-i-1}}{r_\rho}\left[
\frac{\mathfrak d_{j-1}^L\mathfrak d_{N-i}^R}
{\mathfrak d_N^{LR}}
-\mathbf1_{j>i}\mathfrak d_{j-i-1}\right],
\qquad G=(EI-H_{N,\rho}^{v_L,v_R})^{-1}.
\label{eq:Sauxinverse}
\end{align}
To verify it, apply $EI-H_{N,\rho}$ to each column.
Away from the source row the minors obey
Eq.~\eqref{eq:Srecurrence} with the calibrated hoppings;
the subtraction term gives the unit source jump, and the leading
and trailing minors enforce both endpoints.
Normalization cancels the powers of $r_\rho$ exactly.

For both-end control the bounds in Eq.~\eqref{eq:Sauxminorbound}
give
\begin{equation}
|G_{ij}|\le C=\frac{M_1^2/d_*+M_0}{r_*}
=\frac{22045}{840},\qquad
\sigma_{\min}(\mathsf B_N)\ge\frac1{CN}.
\label{eq:Sauxgap}
\end{equation}
The singular-value bound follows from $\|G\|_2\le CN$.
It proves a partner gap of at least $0.0381038784\ldots/N$,
but a sharper estimate is available at the chosen port.
For column $j=1$, the subtraction is absent and $D_0^L=1$:
\begin{equation}
\|\mathsf B_N^{-1}e_1\|_2^2\le NC_p^2,\qquad
C_p=\frac{M_1}{r_*d_*}=\frac{725}{42}.
\label{eq:Sauxcolumn}
\end{equation}
Here $e_1$ is the first unit vector in the $N$-site block.
This first-column estimate controls the closed-state reflection
background without using the larger full inverse norm.

In either one-end setting, the limiting oscillatory amplitude
is at least $\sqrt{11/2}\exp[-B(3/2)^2]$ throughout $I_N^{\rm aux}$.
The bare amplitude is larger. This lower bound equals
$1.6551539838\ldots>1$, leaving a strict amplitude margin.
The half-wave count in Section~\ref{sec:count} therefore gives
$\Theta(\sqrt N)$ simple real roots in this window, up to the
bounded edge contribution, with spacing of order $N^{-1}$.
At such a root $E_*$, the left null vector
$\mathsf B_N(E_*)^Tu=0$ can be normalized by $u_1=1$.
Its transpose recurrence gives
\begin{equation}
u_j=D_{j-1}^L/r_\rho^{j-1}
=(-1)^{j-1}\mathfrak d_{j-1}^L,\qquad
\frac{|u_1|^2}{\|u\|_2^2}\ge\frac1{NM_L^2}.
\label{eq:Sauxportweight}
\end{equation}
The bound uses $M_L=M_0$ for a bare left end and $M_L=M_1$
otherwise. Thus every counted root has weight at least
$1/(NM_0^2)$ at the fixed $A_1$ port of the partner.

\subsection{Reflection contrast and separating valleys}

Choose total amplitude loss $\kappa_N=\varepsilon/N$,
$\varepsilon=0.02$, at every node. At $A_1$, the external port
and internal reservoir each contribute $\kappa_N/2$;
all other nodes have internal loss $\kappa_N$.
For oscillator amplitudes $\mathbf a$ and a unit-flux input
$f_{\rm in}$, the rotating-frame input--output equations are
\begin{align}
\dot{\mathbf a}&=(-i\mathbb K_N-\kappa_N I)\mathbf a
+\sqrt{\kappa_N}\,p f_{\rm in},\nonumber\\
f_{\rm out}&=f_{\rm in}-\sqrt{\kappa_N}\,p^T\mathbf a,
\qquad p=|A_1\rangle .
\label{eq:Sauxio}
\end{align}
The normal drift operator decays at rate $\kappa_N$, with no
transient norm growth. Its stability margin shrinks as $1/N$.

With $\mathcal R_N=(-\mathbb K_N+i\kappa_N I)^{-1}$, the
reflection amplitude $\varrho_N$ and its nominal defect are
\begin{align}
\varrho_N&=1-i\kappa_Np^T\mathcal R_Np,\nonumber\\
q_N=1-\varrho_N
&=\kappa_N^2\left[
(\kappa_N^2I+\mathsf B_N\mathsf B_N^T)^{-1}
\right]_{11}\in[0,1].
\label{eq:Sauxreflection}
\end{align}
Nominal reflection is real; $q_N$ is an amplitude defect,
not a reflected-power deficit.
The $A$ block of $\mathbb K_N^2$ is
$\mathsf B_N\mathsf B_N^T$, with inverse
$\mathsf B_N^{-T}\mathsf B_N^{-1}$.
Its local quadratic form is exactly
$\|\mathsf B_N^{-1}e_1\|_2^2$, and its zero vector is the left
null vector in Eq.~\eqref{eq:Sauxportweight}.

At a real root, the zero term in this positive spectral measure
contributes at least $1/(NM_0^2)$.
In the closed setting, positive matrix inverse ordering and
Eq.~\eqref{eq:Sauxcolumn} give
\begin{align}
q_N^{\rm on}(E_*)&\ge\frac{100}{289N},\nonumber\\
q_N^{\rm off}(E)&\le\frac{\varepsilon^2C_p^2}{N}
=\frac{841}{7056N},\qquad E\in I_N^{\rm aux}.
\label{eq:Sauxsignalbound}
\end{align}
Here the on settings are $(0,0),(r,0),(0,r)$ at their real roots;
off is $(r,r)$ throughout the window.
At the same energy, drive, port and losses the absolute contrast
is at least $0.2268314188\ldots/N$.

The roots also give separated detectable groups.
At phase extrema $\Theta_N=m\pi$, the smallest limiting one-end
minor magnitude is
$(44/125)(1.6551539838\ldots-1)>0.2306$; the bare value is larger.
Thus $|\mathfrak d_N^X|\ge1/5$ there for large $N$.
The first-column formula with trailing-minor bound $M_0$ yields
\begin{equation}
Nq_N(E_{\rm valley})
\le\varepsilon^2\left(\frac{M_0}{r_*/5}\right)^2
=\frac{289}{19600}<\frac14 .
\label{eq:Sauxvalley}
\end{equation}
Adjacent complete phase-extremum intervals have a simple real
root between their subthreshold endpoints.
Their $\Theta(\sqrt N)$ number therefore supplies disjoint
above-threshold groups at $Nq_N=1/4$.
The off setting has no above-threshold point anywhere in the window.
This argument permits additional fine structure inside a group.

\subsection{Precision, waiting time and photons}

The absolute $N^{-1}$ contrast determines the measurement budget.
Retain the declared losses and external coupling and let
$\Delta\mathbb K_N$ be a Hermitian calibration error.
Both damped resolvents have norm at most $1/\kappa_N$;
their exact difference gives
\begin{equation}
|\widetilde\varrho_N-\varrho_N|
\le\frac{\|\Delta\mathbb K_N\|_2}{\kappa_N}.
\label{eq:Sauxcalibration}
\end{equation}
A scaled calibration allocation $b_{\rm cal}=0.02$ is met by
$\|\Delta\mathbb K_N\|_2\le0.0004/N^2$.
For errors on the at most five links per row, an absolute error
$0.00008/N^2$ per link suffices.
The reflection estimate allows chirality-breaking Hermitian
errors, although exact zero crossings refer to the nominal
chiral network. Loss and port-coupling errors need separate bounds.

At unit incident flux, a steady state has amplitude norm at most
$1/\sqrt{\kappa_N}$. A change between two steady-state settings
leaves an output transient bounded by $2e^{-\kappa_Nt}$.
For a scaled settling allocation $b_{\rm set}=0.02$, wait
\begin{equation}
t_{\rm set}\ge\kappa_N^{-1}\log(2N/b_{\rm set})
=O(N\log N).
\label{eq:Sauxsettling}
\end{equation}
A vacuum restart replaces the factor $2N$ by $N$.
The logarithm is needed because the allowed output error
shrinks as $1/N$.

For ideal coherent homodyne detection with vacuum reservoirs,
$M$ incident photons per point give amplitude standard deviation
$1/(2\sqrt M)$. Write $\mu_j$ for the actual mean real quadrature
after calibration and settling, and $\widehat\varrho_j$ for its
estimator. Allocate $b_{\rm hom}=0.05$ and let $\mathcal J_N$
be the total sampled points across all four scans.
A Gaussian union bound gives
\begin{align}
\Pr\!\left(\max_j|\widehat\varrho_j-\mu_j|>
b_{\rm hom}/N\right)
&\le2\mathcal J_N e^{-2Mb_{\rm hom}^2/N^2},\nonumber\\
M&\ge\frac{N^2}{2b_{\rm hom}^2}
\log\frac{2\mathcal J_N}{p_{\rm fail}}.
\label{eq:Sauxphotons}
\end{align}
The chosen failure probability $p_{\rm fail}$ covers all four
scans. Multiple sizes require combining point counts or failure
budgets; independence between points is unnecessary.
Calibration and settling bound the mean offsets separately.
Consequently, with $\widehat q=1-\widehat\varrho$,
the total nominal amplitude error is at most
$b_{\rm tot}/N$, where
\begin{equation}
b_{\rm tot}=b_{\rm hom}+b_{\rm set}+b_{\rm cal}=0.09.
\label{eq:Sauxerrorbudget}
\end{equation}
Thermal noise, detection inefficiency and technical noise require
an experimental noise model beyond this ideal vacuum budget.

Let $L$ be a size-independent bound on $\|\mathbb K_N'(E)\|_2$.
The resolvent identity yields $|q_N'|\le2L/\kappa_N$.
A blind grid of spacing
\begin{equation}
h_N\le\frac{0.003\,\varepsilon}{LN^2}
\label{eq:Sauxblindgrid}
\end{equation}
loses at most $0.003/N$ at the nearest sample to a root or valley.
Since $100/289-0.09-0.003>1/4$, the noisy sampled on responses
retain separated above-threshold groups; the off and valley
upper bounds remain below $1/4$.
There are $\mathcal J_N=O(N^{3/2})$ settings across the shrinking
window and four scans. At fixed $p_{\rm fail}$,
\begin{equation}
M\mathcal J_N=O(N^{7/2}\log N),\qquad
t_{\rm set}\mathcal J_N=O(N^{5/2}\log N).
\label{eq:Sauxresources}
\end{equation}
At fixed incident flux, photon integration takes the longer time.
The construction uses $O(N)$ modes, bounded couplings and
polynomial accuracy and measurement resources.
It is a passive readout in an independent network, with both
signal and stability margin shrinking as $1/N$; it establishes
no sensing advantage in the original device or over an optimized
Hermitian sensor \cite{LauClerk2018Sensing}.

\subsection{What the finite sampled profiles certify}
\label{sec:auxprofile}

Figure~4 uses the faster grid $0.01/N$.
Its finite-size count is checked directly; the general blind-grid
bound above uses $O(N^{-2})$ and has no certified onset size.
For a sampled profile put $y_j=Nq_N(E_j)$ and
$\widehat y_j=N\widehat q_N(E_j)$, with
$|\widehat y_j-y_j|\le\delta_{\rm prof}=0.09$.
The support/seed rule accepts a connected group above
$h_{\rm sup}=0.20$ only if its maximum exceeds
$h_{\rm seed}=0.40$.
Define reference cores by
\begin{equation}
y_j>h_{\rm sup}+\delta_{\rm prof}=0.29.
\label{eq:Sauxprofilecore}
\end{equation}
If every core has maximum $>0.49$ and adjacent cores have a
separating point with $y_j\le0.11$, each core remains connected
above support, retains a seed, and cannot bridge to its neighbor.
Outside a core $\widehat y_j\le0.38<0.40$, so no new seed forms.
Accepted noisy groups therefore correspond one-to-one to these
reference cores under the stated pointwise error bound.

Within a core of nominal maximum $y_{\max}$, any candidate noisy
maximum must have $y_j\ge y_{\max}-0.18$.
The scan extends by $1/N$ beyond each counting-window endpoint.
A core is inside only if all candidate maximum positions are inside;
it is outside only if all are strictly on the same exterior side.
Candidates straddling an endpoint remain unresolved.
Components cut by an outer scan endpoint and root-to-core matching
are checked separately. The analytic $1/4$ separation threshold
does not imply the $0.49$ seed condition of this finite-profile
certificate.

Real diagonal errors $\xi_j$ in $\mathsf B_N$ have a more
structured tolerance: they become inter-sublattice rung errors.
For sufficiently small fixed $\sum_j|\xi_j|$, the transfer-product
argument in Section~\ref{sec:spatialerror} preserves bounded leading
and trailing minors and a positive two-end determinant margin.
The left-null recurrence and first-column formula remain exact
with the same clean $\rho(E)$, so the signal estimates persist
with relaxed constants and no sample-dependent bulk retuning.
This total-error result concerns those rung errors; generic
coupling errors use Eq.~\eqref{eq:Sauxcalibration}.

\section{Complex branches, exceptional points and finite resolution}
\label{sec:complex}

The same determinant supplies the asymptotic curves in Fig.~2(a)
of the Letter and the resolution scale discussed in its final section.
Let $\xi_{k,N}$ be a real phase center,
$\Theta_N(\xi_{k,N},x/N)=(2k+1)\pi$, and put
$s=\sqrt N(\xi_{k,N}-\Ec)$, $z=N(E-\xi_{k,N})$.
Local holomorphic continuation gives, on compact $s,x,z$ sets,
\begin{equation}
F_N(\xi_{k,N}+z/N,x/N)
=1-\mathfrak a(s,x)\cos(bz)+O(N^{-1/2}),\qquad
\mathfrak a=Q_ce^{-Ax-Bs^2}.
\label{eq:Sholo}
\end{equation}
The error follows from $Ne^3=O(N^{-1/2})$, the cell displacement
$O(N^{-1})$ and the subdominant gap; Cauchy estimates control
derivatives on interior compact sets. Away from $\mathfrak a=1$,
the limiting roots are simple:
\begin{equation}
z_\pm=
\begin{cases}
\pm b^{-1}\arccos(\mathfrak a^{-1}),&\mathfrak a>1,\\
\pm i b^{-1}\operatorname{arcosh}(\mathfrak a^{-1}),&0<\mathfrak a<1.
\end{cases}
\end{equation}
Rouch\'e's theorem on disjoint small circles gives one finite-$N$
root near each limiting root. Its nonzero derivative sets the
energy displacement to $O(N^{-3/2})$. Conjugation symmetry and
uniqueness force the real-branch roots to be exactly real.
For a complex pair,
\begin{equation}
N|\operatorname{Im}E|
=\frac1b\operatorname{arcosh}\!\left(\frac{e^{Ax+Bs^2}}{Q_c}\right)
+O(N^{-1/2}).
\label{eq:Spole}
\end{equation}
Replacing the center's $s$ by the pole's scaled real part keeps
this error order. These curves are derived, not fitted.
The simple-root displacement is not uniform at the double-root
limit, although Eq.~\eqref{eq:Sholo} remains uniform.

\subsection{Finite-matrix exceptional points}
\label{sec:ep}

Near the last surviving half-wave, the envelope maximum is
$\log Q_c-AN\dd+O(N^{-1})$, at $e=O(N^{-1})$.
The nearest phase center changes it by $O(N^{-1})$, while its
detuning derivative is $-AN+O(1)$. The final tangency thus occurs at
\begin{equation}
\dd_{\rm EP,last}=\frac{\log Q_c}{AN}+O(N^{-2}),\qquad
E_{\rm EP,last}-\Ec=O(N^{-1}).
\label{eq:Sep}
\end{equation}
The correction depends on phase and need not be monotone.
At a tangency, $F_N=F_{N,E}=0$, and
\begin{equation}
F_{N,EE}=b^2N^2[1+o(1)]\ne0,\qquad
F_{N,\dd}=AN[1+o(1)]\ne0 .
\label{eq:Sderivativesep}
\end{equation}
Hence
\begin{equation}
(E-E_{\rm EP})^2
=-\frac{2F_{N,\dd}}{F_{N,EE}}(\dd-\dd_{\rm EP})
+o(|\dd-\dd_{\rm EP}|).
\label{eq:Spuiseux}
\end{equation}
Algebraic multiplicity is exactly two. Geometric multiplicity is
one: the nonzero lower subdiagonal $r$ lets the last row determine
$u_{N-1}$ from $u_N$, and successive rows determine every preceding
component. If $u_N=0$ the vector vanishes. Each eigenspace is
therefore one-dimensional, so these double zeros are EP2s.
The bulk spatial roots remain distinct.

\begin{figure}[t]
\includegraphics[width=.76\textwidth]{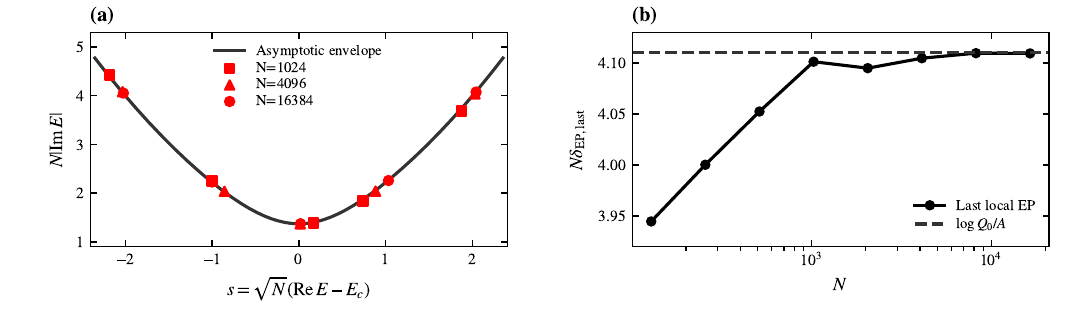}
\caption{Complex branches and finite-matrix exceptional points.
(a) At $v_L=v_R=1.5$, $\dd=0$, the scaled real part
$\sqrt N(\operatorname{Re}E-\Ec)$ and scaled imaginary part
$N|\operatorname{Im}E|$ are compared with Eq.~\eqref{eq:Spole}.
(b) Last local EP for bare endpoints. The dashed detuning limit is
$\log Q_0/A=4.110381306$. Phase-dependent corrections in
Eq.~\eqref{eq:Sep} need not be monotone; the bulk roots do not coalesce.}
\label{fig:Spoles}
\end{figure}

\subsection{What finite imaginary-energy resolution counts}
\label{sec:resolution}
\label{thm:resolution}

For fixed $S>0$, $x$ and $y\ge0$, define the local count
\begin{equation}
N_{y/N}(S,x)=\#\{E_j:\,
|\sqrt N(\operatorname{Re}E_j-\Ec)|<S,\quad
|\operatorname{Im}E_j|\le y/N\},
\qquad \dd=x/N.
\label{eq:Sresolutiondef}
\end{equation}
Multiplicity is included. The non-strict imaginary condition
makes $y=0$ the strict-real count in this shrinking window.

To exclude extra roots, use a phase-cell rectangle with vertical
sides at the adjacent real solutions of $\Theta_N=2k\pi$ and
$2(k+1)\pi$, and horizontal sides at $\operatorname{Im}E=\pm Y/N$.
Choose a fixed $Y>y$ exceeding all limiting root heights over
$|s|\le S$. In the $z$ coordinate the vertical sides approach
$\pm\pi/b$, where the limiting function is
$1+\mathfrak a\cosh(b\operatorname{Im}z)>0$.
There are no roots on the horizontal sides either. The boundary
modulus has a uniform positive lower bound, so Rouch\'e's theorem
gives exactly two roots per cell, including multiplicity at
$\mathfrak a=1$. Adjacent cells have disjoint interiors and cover
the strip, with only $O(1)$ cells cut by the real-window edges.

On the closed side the limiting height is
$b^{-1}\operatorname{arcosh}(e^{Ax+Bs^2}/Q_c)$; on the real side
it is zero. Height at most $y$ is equivalent to $Bs^2\le K_y$,
where $K_y=\log Q_c-Ax+\log\cosh(by)$.
The center density in $s$ is $b\sqrt N/(2\pi)$.
Small neighborhoods of the height cutoff and $\mathfrak a=1$
contain $O(\varepsilon\sqrt N)+O(1)$ cells; each still has two roots.
Counting outside these neighborhoods, then taking $N\to\infty$
and $\varepsilon\to0$, yields
\begin{equation}
\lim_{N\to\infty}\frac{N_{y/N}(S,x)}{\sqrt N}
=\frac{2b}{\pi}\min\left\{S,\sqrt{\frac{[K_y]_+}{B}}\right\},
\qquad [t]_+=\max(t,0).
\label{eq:Sresolution}
\end{equation}
At $y=0$, the exact half-wave count in Section~\ref{sec:count}
gives the same result. If $K_y<0$, all limiting heights exceed
$y$ and the count is eventually zero. At $K_y=0$ only the
normalized limit is specified; no universal finite integer or
uniform $O(1)$ remainder follows.

\begin{figure}[t]
\includegraphics[width=.76\textwidth]{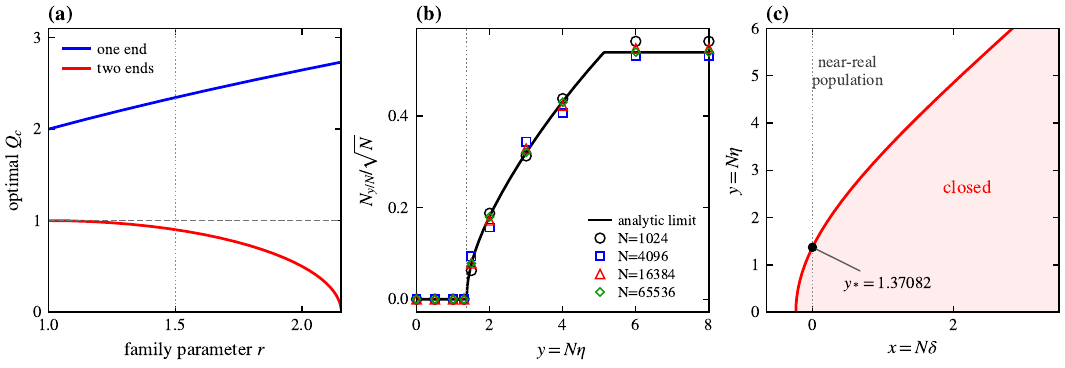}
\caption{Endpoint control and finite resolution of the original
non-Hermitian spectrum. (a) Minimum one-end and two-end coefficient
ratios in the open family; the dashed line is the threshold at $x=0$.
(b) Near-real counts for $r=3/2$, $(v_L,v_R)=(1.5,1.5)$,
$x=0$, and scaled real-window half-width $S=2.5$.
Symbols enumerate finite-determinant roots by phase cell;
the curve is Eq.~\eqref{eq:Sresolution}.
(c) The closing boundary $Q_c\cosh(by)=e^{Ax}$ for the same model
and optimal double-end potentials as in (b), with scaled detuning
$x=N\dd$ and resolution $y=N\eta$ varied.
Equality has no universal finite count.}
\label{fig:Sauxresolution}
\end{figure}

Thus the resolution-dependent closing condition is
\begin{equation}
Q_c\cosh(by)<e^{Ax}.
\label{eq:Sresolutionclosure}
\end{equation}
For optimal double-end control at $x=0$,
$y_*=b^{-1}\operatorname{arcosh}(4/\sqrt{13})=1.37082051259\ldots$.
Fixed $y<y_*$ preserves closure; fixed $y>y_*$ gives a
square-root-order near-real population. This is the fixed-$y$
limit $\eta=y/N$, not an estimate for fixed absolute $\eta$.
Figure~\ref{fig:Sauxresolution} shows the original spectral count,
separately from the passive reflection measurement.

\section{Spatial errors without retuning the bulk}
\label{sec:spatialerror}
\label{thm:spatialerror}

The final Discussion claim concerns the representative chain at
$\dd=0$, with the four fixed boundary settings, under arbitrary
real on-site errors. Define
\begin{equation}
H_N^{v_L,v_R,\xi}=H_N^{v_L,v_R}
+\sum_{j=1}^N\xi_j|j\rangle\langle j|,\qquad
\|\xi\|_1=\sum_j|\xi_j|,\qquad
I_N(S)=\{|E-\Ec|<S/\sqrt N\},\quad S=5/2.
\label{eq:Sspatialmodel}
\end{equation}
Positions and signs are arbitrary and may depend on $N$.
For a sufficiently small size-independent total budget $\tau$
with $\|\xi\|_1\le\tau$, the double-end setting has no real root in
$I_N(S)$, whereas the other three settings satisfy
\begin{equation}
c_-\sqrt N\le N_{\mathbb R}^{\xi}(I_N(S))\le c_+\sqrt N
\label{eq:Sspatialcount}
\end{equation}
for all sufficiently large $N$, with $c_\pm>0$ independent of
the error locations. The count includes multiplicity.
Neither $\Ec$ nor $\dd$ is retuned. The clean leading coefficient
and its $O(1)$ counting remainder are not retained by this bound.

For the leading $n$ sites let $d_n^\xi$ be their determinant,
with $d_0^\xi=1$ and $d_n^\xi=0$ for $n<0$.
The exact inhomogeneous recurrence is
\begin{equation}
d_n^\xi=(E-h_n)d_{n-1}^\xi-r t_1d_{n-2}^\xi
-r^2t_2d_{n-3}^\xi-r^3t_3d_{n-4}^\xi,\qquad
h_n=\xi_n+v_L\delta_{n1}+v_R\delta_{nN}.
\label{eq:Sspatialrecurrence}
\end{equation}
Here $\delta_{nm}$ is the Kronecker delta.
Normalize by $\alpha_A(E)^n$, with $\alpha_A=r/\beta_0$.
The clean companion matrix has distinct eigenvalues
$1,\beta_0/\beta_+,\beta_0/\beta_-,\beta_0/\beta_3$ of modulus
at most one on a small real neighborhood, because
$\Delta(E,0)\le0$. Its diagonalizing matrix $V(E)$ and inverse
are uniformly bounded. In the norm induced by $V$, the clean
transfer has norm at most one; an error changes its first entry
by $-\xi_n/\alpha_A$, with norm at most $C|\xi_n|$.

The fixed endpoint transfers are bounded in the same basis.
Telescoping the product, including endpoint errors, gives a
change bounded by $C(e^{C\|\xi\|_1}-1)$.
The clean real coefficient
$c_A^{v_L,v_R}=c_A(1-v_L/\alpha_A)(1-v_R/\alpha_A)$ and
$\alpha_A$ remain nonzero in a fixed complex neighborhood.
After normalization,
\begin{equation}
\sup_{I_N(S)}|F_N^\xi-F_N^0|
\le C_1(e^{C_2\|\xi\|_1}-1),\qquad
F_N^\xi=\frac{d_N^\xi}{c_A^{v_L,v_R}\alpha_A^N}.
\label{eq:Sspatialbound}
\end{equation}
The constants can be chosen for all four settings.
For both ends, $F_N^0\ge1-\sqrt{13}/4-o(1)$ throughout
$I_N(S)$. A small common $\tau$ preserves this positive margin.
For the other settings, use $S_0=3/2<S$.
Their smallest clean amplitude is
$\sqrt{11/2}e^{-BS_0^2}>1$. At the alternating phase points
$\Theta_N(E_k)=k\pi$, the clean determinants have opposite signs
with uniform margins, and the phase advances by
$2bS_0\sqrt N+O(1)$ across the inner window.
The same $\tau$ preserves these signs, giving at least
$(2bS_0/\pi)\sqrt N-O(1)$ real roots by continuity.
For example, let
$\mu=\frac14\min\{1-\sqrt{13}/4,\sqrt{11/2}e^{-BS_0^2}-1,1\}>0$,
take $\tau=C_2^{-1}\log(1+\mu/C_1)$, and then choose the onset
size so the clean margins exceed $2\mu$.

An amplitude bound alone cannot rule out extra roots.
Continue $\alpha_A,V,c_A$ and the nonzero normalized transfer
eigenvalues analytically into the fixed complex neighborhood.
The matrices $V,V^{-1}$ and the coefficients remain uniformly bounded.
The transfer eigenvalues have bounded logarithmic derivatives, so
their moduli are at most $e^{C|\operatorname{Im}E|}$.
The product estimate becomes
$|F_N^\xi(E)|\le C e^{CN|\operatorname{Im}E|+C\|\xi\|_1}$.
An even phase point $E_N=\Ec+O(N^{-1})$ has
$|F_N^\xi(E_N)|$ bounded away from zero.
Jensen's formula on the disk of radius $4S/\sqrt N$ about
$E_N$ bounds the zeros in the concentric $2S/\sqrt N$ disk
by $O(\sqrt N)$, including multiplicity.
The inner disk contains $I_N(S)$ for large $N$, giving the upper
bound in Eq.~\eqref{eq:Sspatialcount}.

\begin{figure}[t]
\includegraphics[width=.76\textwidth]{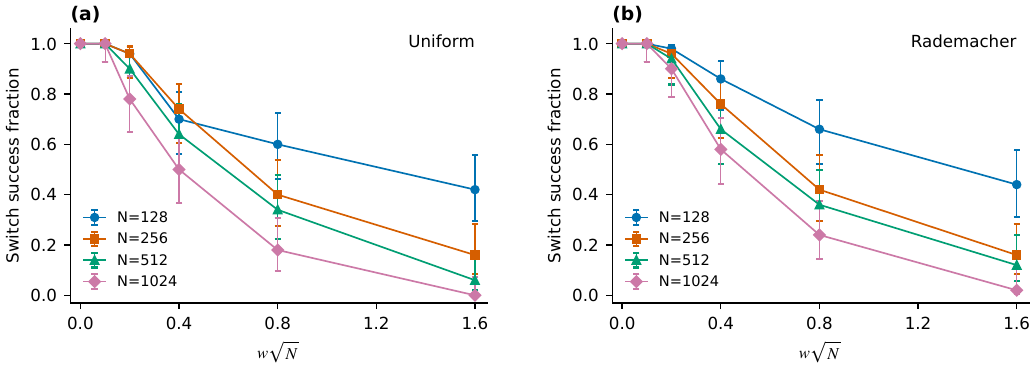}
\caption{Finite random on-site errors at the nominal bulk contact.
Errors are $\xi_j=wX_j$, with zero-mean unit-variance
(a) uniform and (b) two-point $X_j$; $N$ is chain length.
A realization is successful when the two-end count is zero and
each bare or single-end count retains at least half its clean value
in $|E-\Ec|<(5/2)/\sqrt N$.
Each nonzero-strength point uses 50 realizations shared by all
four settings and a pointwise 95\% Wilson interval.
Zero strength is a deterministic baseline. Lines connect samples;
the size dependence does not establish a universal collapse.}
\label{fig:Sdisorder}
\end{figure}

This total-error condition permits finitely many sufficiently weak
impurities at arbitrary sites; when every site is affected,
worst-case $O(N^{-1})$ amplitude suffices. It does not optimize
single-end controls in a disordered sample or cover fixed-strength
extensive disorder. The finite random-error tests in
Fig.~\ref{fig:Sdisorder} examine a different, sampled question.

\section{Numerical methods for the Letter's figures}
\label{sec:numerics}

\emph{Spectra in Fig.~2.}
At $N=256$, the 36 displayed local eigenvalues are compared
between direct diagonalization and a diagonal similarity with
hoppings $t_n(1.03)^n$. Their maximum displacement is
$3.19\times10^{-13}$ and maximum relative residual in
Eq.~\eqref{eq:Sboundary} is $1.18\times10^{-11}$.
The asymptotic curves use Eq.~\eqref{eq:Spole} and its real branch.
The phase diagram evaluates Eq.~\eqref{eq:SQ} directly.

\emph{Counts in Fig.~3.}
The original long-chain counts evaluate the exact normalized determinant
on a sign-change grid and refine each bracket by Brent's method.
The half-width is
$\epsilon_N=\min\{0.43,\,1.5\sqrt{([K]_++3)/(BN)}\}$,
where $K=\log Q_c-Ax$ and $[K]_+=\max(K,0)$.
The grid has $\max\{65,\lfloor2\epsilon_N/h\rfloor+1\}$ points,
with nominal step $h=\pi/(0.5N\times16)$; the root solver uses
absolute tolerance $5\times10^{-15}$ and relative tolerance $10^{-14}$.
Later independent counts bracket roots between derivative extrema.
The independent leading-minor recurrence is
\begin{equation}
D_n=E D_{n-1}-r t_1D_{n-2}-r^2t_2D_{n-3}-r^3t_3D_{n-4},
\qquad D_0=1,\quad D_{-1}=D_{-2}=D_{-3}=0.
\label{eq:Srecurrence}
\end{equation}
At $x=0$, the bare counts for
$N=128,256,512,1024,2048,4096,8192,16384$ are
$8,12,16,24,32,48,66,94$; optimal double-end counts are zero.
These numerical windows shrink with size and enclose the predicted
population; Eq.~\eqref{eq:Scount} instead counts in a fixed small interval.
Threshold and singular counts use the
smooth unnormalized determinant at 24 and 48 samples per phase scale,
with agreeing counts. Selected roots are refined by the scaled
recurrence at 80 decimal digits. EPs solve both determinant equations
at 65 digits.

The resolution calculations classify each phase cell by the real-axis
derivative minimum before real bracketing or complex-pair solving.
They use $|\sqrt N(\operatorname{Re}E-\Ec)|<2.5$, which can
truncate the full population. Selected $N=65536$ roots are checked
at 80 digits; the largest recurrence residual is
$1.49\times10^{-76}$.
Eighteen rational-polynomial cases at $N=64,128,256$ have exact
real-root isolation in $[\Ec-1/4,\Ec+1/4]$, and eight independent
Sturm counts agree. These finite intervals are certified; the
other long-chain counts and complex scans are numerical checks.

\emph{Reflection in Fig.~4.}
The same four settings use $\kappa_N=0.02/N$ and a blind grid
$0.01/N$ in $|\sqrt N(E-\Ec)|<1.5$, extended by $1/N$
at both endpoints. The banded solve for Eq.~\eqref{eq:Sauxreflection}
is compared with dense inverses, singular-value decompositions and
input--output energy balance on smaller cases, including five
high-precision checks. Independent complete-window root enumeration
and derivative-extremum counts are matched to the profile cores.
Across $N=64$ through $8192$, the finite-profile conditions in
Section~\ref{sec:auxprofile} certify the displayed sampled counts.
At $N=8192$ they are $30,29,29,0$.
This profile certificate is distinct from rational-polynomial
root certification and supplies no certified onset size for the
asymptotic signal bounds.

\emph{Spatial-error checks.}
For Fig.~\ref{fig:Sdisorder}, independent $X_j$ are uniform on
$[-\sqrt3,\sqrt3]$ or equally likely $\pm1$; every realization is
reused across the four settings and across strengths.
No sample-dependent bulk retuning is applied.
The scaled window is $S=5/2$; the numerical strict-real cutoff
is $|\operatorname{Im}E|\le10^{-8}$.
Near-real checks separately use $y/N$, $y=0.5,1,1.5,2$.
The extension uses $N=128,256,512,1024$ and 50 realizations
per size, strength and distribution.
At $w\sqrt N=0.1$, each tested set has 50 successes;
at $N=1024$, $w\sqrt N=0.2$, the two distributions give
39 and 45, and at $0.4$ they give 25 and 29.
Disputed locations are checked with the inhomogeneous recurrence
at two grid densities and, where needed, 80 digits.
Twenty-four rational nonuniform matrices at $N=32,64,96$
have exact Sturm counts in $\Ec\pm1/4$, including restored
two-end roots. These are finite examples, separate from the
small-total-error argument above.

\end{document}